\documentclass[aps,prb,reprint,floatfix,superscriptaddress]{revtex4-2}

\usepackage{amsmath,amsfonts,amssymb}
\usepackage{graphicx}
\usepackage{dcolumn}
\usepackage{bm}
\usepackage{braket}
\usepackage{xcolor}
\usepackage{tikz}
\usepackage{ulem}
\usepackage[caption=false]{subfig}
 \usepackage[version=4]{mhchem}
\usetikzlibrary{quantikz2}
\usepackage{multirow}
\usepackage{makecell}
\usepackage{amsthm}
\usepackage{soul}
\usepackage{appendix}
\usepackage{hyperref}
\usepackage{subfig}
\usepackage{tabularx}
\usepackage{booktabs}
\usepackage{array}

\newcommand{\bea}{\begin{eqnarray}}
\newcommand{\eea}{\end{eqnarray}}
\newcommand{\eq}[1]{Eq.~(\ref{#1})} %
\newcommand{\fig}[1]{Fig.~\ref{#1}} %

\newcommand{\Var}{\text{Var}}

\newcommand{\imag}{\textrm{i}}

\begin{document}


\title{Quantum Seniority-based Subspace Expansion: 
Linear Combinations of Short-Circuit Unitary Transformations for the
Electronic Structure Problem} 

\author{Smik Patel}
\affiliation{Chemical Physics Theory Group, Department of Chemistry, University of Toronto, Toronto, Ontario M5S 3H6, Canada}
\affiliation{Department of Physical and Environmental Sciences, University of Toronto Scarborough, Toronto, Ontario M1C 1A4, Canada }

\author{Praveen Jayakumar}
\affiliation{Chemical Physics Theory Group, Department of Chemistry, University of Toronto, Toronto, Ontario M5S 3H6, Canada}
\affiliation{Department of Physical and Environmental Sciences, University of Toronto Scarborough, Toronto, Ontario M1C 1A4, Canada }

\author{Rick Huang}
\affiliation{Chemical Physics Theory Group, Department of Chemistry, University of Toronto, Toronto, Ontario M5S 3H6, Canada}

\author{Tao Zeng}
\affiliation{Department of Chemistry, York University, Toronto,  Ontario M3J 1P3, Canada}

\author{Artur F. Izmaylov}%
\email{artur.izmaylov@utoronto.ca}
\affiliation{Chemical Physics Theory Group, Department of Chemistry, University of Toronto, Toronto, Ontario M5S 3H6, Canada} \affiliation{Department of Physical and Environmental Sciences, University of Toronto Scarborough, Toronto, Ontario M1C 1A4, Canada }

\date{\today}

\begin{abstract}

Quantum SENiority-based Subspace Expansion (Q-SENSE) is a hybrid quantum-classical algorithm that interpolates between the Variational Quantum Eigensolver (VQE) and Configuration Interaction (CI) methods. It constructs Hamiltonian matrix elements on a quantum device and solves the resulting eigenvalue problem classically. Unlike other expansion-based methods — such as Quantum Subspace Expansion (QSE), Quantum Krylov Algorithms, and the Non-Orthogonal Quantum Eigensolver — Q-SENSE introduces seniority operators as artificial symmetries to construct orthogonal basis states. This seniority-symmetry-based approach reduces one of the primary limitations of VQE on near-term quantum hardware — circuit depth — at the cost of measuring additional matrix elements. The artificial symmetries also reduce the number of Hamiltonian terms that must be measured, as only a small fraction of the terms couple basis states in different seniority subspaces. With all these merits, Q-SENSE offers a scalable and resource-efficient route to quantum advantage on near-term quantum devices and in the early fault-tolerant regime.

\end{abstract}

\maketitle

\section{Introduction}\label{Section:Introduction}
Understanding the low-energy spectrum of the electronic Hamiltonian lies at the heart of quantum chemistry. An important question is whether quantum computers can provide a practical advantage for solving the electronic structure problem before large-scale quantum error correction becomes available. Early hopes for quantum advantage without error correction were inspired by the development of the Variational Quantum Eigensolver (VQE) \cite{vqe2014}, a hybrid algorithm designed to avoid the deep circuits required by Quantum Phase Estimation (QPE) \cite{Abrams1999QPE,Aspuru‑Guzik2005}. However, despite its promise, VQE faces significant scalability challenges. As system size increases, achieving chemical accuracy requires deeper and more complex circuits to represent the ground-state wavefunction. This growing circuit depth eventually exceeds the coherence capabilities of near-term hardware, making VQE unsuitable for large-scale quantum chemistry without error correction. In addition, VQE faces two further major limitations: (1) the quantum measurement problem, which arises from the need to evaluate a large number of Pauli expectation values in the Hamiltonian \cite{patelQuantumMeasurementQuantum2025}, and (2) the non-linear optimization problem, which involves tuning parameters of unitary operators over a rugged cost landscape. While physically motivated ansätze such as qubit- or unitary-coupled-cluster (QCC/UCC) alleviate some of the optimization difficulty by using energy gradients, the underlying limitations remain \cite{vqe2014,kandalaHardwareefficientVariationalQuantum2017,ryabinkinQubitCoupledCluster2018,grimsleyAdaptiveVariationalAlgorithm2019a,tangQubitADAPTVQEAdaptiveAlgorithm2021a,anandQuantumComputingView2022,anastasiouTETRISADAPTVQEAdaptiveAlgorithm2024}. An attempt to reduce the circuit cost using hardware-efficient unitaries usually leads to barren plateaus~\cite{BP_McClean2018,BP_Cerezo2021}. 

A natural solution to the circuit depth bottleneck is to introduce flexibility in the wavefunction ansatz: rather than representing the eigenstate as the output of a single, deep quantum circuit, one can instead use a linear combination of states obtained using shallower circuits:
\bea\label{eq:qse}
\ket{\Psi} = \sum_{\mu} c_\mu \hat{U}_\mu \ket{\rm HF},
\eea
where $\ket{\rm HF}$ is the Hartree-Fock reference state, $\hat{U}_\mu$ are shallow unitary circuits, and $c_\mu$ are classical coefficients. Finding $c_\mu$ for given basis states $\ket{\phi_\mu} = \hat{U}_\mu\ket{\rm HF}$ is done by solving the generalized eigenvalue problem on a classical computer 
\begin{align}
    \mathsf{H}\vec{c} &= E\mathsf{S}\vec{c}\label{GEVP}\\
    \mathsf{H}_{\mu\nu} &= \braket{\phi_\mu|\hat{H}|\phi_\nu}\label{matrix_elements}\\
    \mathsf{S}_{\mu\nu} &= \braket{\phi_\mu|\phi_\nu}\label{overlaps}
\end{align}
after obtaining matrix elements $\mathsf{H}_{\mu\nu}$ and $\mathsf{S}_{\mu\nu}$ through measurements on a quantum computer. This framework is known as Quantum Subspace Expansion (QSE), and a comprehensive review of it is provided in Ref.~\cite{mottaSubspaceMethodsElectronic2024}.
 
Interestingly, early developments of expansion-based methods, such as Quantum Subspace Expansion VQE (QSE-VQE)~\cite{McClean2017} and Multistate Contracted VQE (MC-VQE)~\cite{Parrish2019}, did not have circuit reduction as their motivation; instead they aimed to obtain excited states. In 2020, Huggins {\it et al.}~\cite{NOVQE2020} proposed using smaller circuits for preparing $\ket{\phi_\mu}$ to reduce the VQE circuit cost. The remaining problems were reducing the number of measurements and optimizing the unitaries $\hat{U}_\mu$. These two problems arise from the non-orthogonality of the basis states, which can lead to ill-conditioned overlap matrices $\mathsf{S}$, implying the need for more measurements to reach a desired accuracy~\cite{Lee2025}.

To address the optimization problem in QSE, Krylov-based \cite{parrishQuantumFilterDiagonalization2019} and Variational QPE~\cite{Klemenko2022} approaches were proposed. These methods use unitaries $\hat{U}_\mu = e^{-\imag \hat{H}t_\mu}$, where $\hat{H}$ is the full electronic Hamiltonian. Due to the complexity of $\hat{H}$, these unitaries require Trotter approximations~\cite{TrotterRef} or qubitization~\cite{QubitizationRef}, both of which have significant resource demands. Another optimization approach is the Non-Orthogonal Quantum Eigensolver (NOQE)~\cite{Baek2023}, which uses UCC singles and doubles generators for the $\hat{U}_\mu$ and estimates their amplitudes using M{\o}ller-Plesset perturbation theory. Despite reducing circuit complexity and avoiding the need for Trotterization, NOQE still faces challenges in measuring matrix elements. Recently, Ren \textit{et al.} proposed the use of classical shadow tomography \cite{CST_Huang2020} to address this challenge~\cite{NOQE_CST2025}. Although this reduces the scaling of the measurement problem with respect to the number of $\ket{\phi_\mu}$ states from quadratic to linear, the classical post-processing cost may grow substantially.  

The choice of the $\hat{U}_\mu$ is one of the key issues in many QSE methods. Ideally, one would like to have unitaries $\hat{U}_\mu$ that ensure: 1) circuit-depth adaptability depending on quantum hardware, 2) orthogonality, 3) the possibility of optimization using both classical and quantum methods, and 4) efficient measurement of the Hamiltonian matrix elements. 

In this work, we introduce Quantum SENiority-based Subspace Expansion (Q-SENSE). The key idea of Q-SENSE is to build the $\hat{U}_\mu$ so that the basis states $\ket{\phi_\mu}$ are eigenstates of the seniority operator~\cite{seniority2011}, which counts the number of unpaired electrons in the wavefunction. Based on empirical evidence, this operator provides a systematically improvable approach for modeling strongly correlated systems by constructing the wavefunction approximation with components of increasing seniority. Already, the zero seniority sector provides a qualitatively correct dissociation of multiple bonds in molecules~\cite{shepherd2016fciqmc,boguslawski2019benchmarking}. 
As with any Hermitian operator, eigenstates of different seniority values are orthogonal. This allows unconstrained optimization of the parameters of $\hat{U}_\mu$ across different seniority sectors without risk of generating non-orthogonal basis states.

Unlike a zero-seniority-only approach~\cite{elfving2020seniority}, Q-SENSE systematically includes higher-seniority sectors. Yet, each basis state has a distinct set of singly and doubly occupied orbitals. Thus, we can benefit from compressed qubit encodings, generalizing the Hard-Core Boson encoding used for seniority-zero wavefunctions in Refs.~\cite{elfvingSimulatingQuantumChemistry2021,kottmannOptimizedLowdepthQuantum2022}. Although using this hybrid qubit encoding for the total eigenfunction $\ket{\Psi}$ in \eq{eq:qse} would introduce an approximation, this is not the case here when using the encoding for each basis state $\ket{\phi_\mu}$.

Finally, although our approach may appear limited to the pre-fault-tolerant era, it is equally applicable in the early fault-tolerant regime. Just as VQE wavefunctions can serve as inputs to QPE for improved eigenvalue estimation, Q-SENSE wavefunctions—being linear combinations of unitaries—can similarly be used for state preparation in QPE pipelines.

\section{Q-SENSE Algorithm}

In this section, we discuss the definition of the Q-SENSE basis states and the estimation of their Hamiltonian matrix elements.

\subsection{Q-SENSE Basis}

To construct Q-SENSE basis states preserving seniority symmetry we apply parametrized unitaries on the Hartree-Fock state, $\ket{\rm HF}$, with $N_e$ electrons. We will use $a,b,c,\ldots$ ($i,j,k,\ldots$) to index virtual (occupied) orbitals of $\ket{\rm HF}$ and $r,s,t,\ldots$ for all orbitals. The basis states take the following form   
\bea 
\ket{\phi_\mu} &=& \hat{U}_\mu\ket{\rm HF} = \hat{W}_\mu \hat{V}_\mu \ket{\rm HF}, 
\eea
where $\hat{W}_\mu$ are electron-pair rotations  
\bea
\hat{W}_\mu (\theta) &=& \prod_{r,s} \exp\Big(\theta_{rs}\hat{T}_{rs}^{(p)}\Big),\label{eq:W} \\
\hat{T}_{rs}^{(p)} &=& \hat{a}_{r\uparrow}^\dagger \hat{a}_{r\downarrow}^\dagger \hat{a}_{s\downarrow} \hat{a}_{s\uparrow} - h.c. \label{pair_excitation}
\eea
and $\hat{V}_\mu$ are unitaries that create configuration state functions (CSFs) from $\ket{\rm HF}$, $\ket{\rm CSF_\mu} = \hat{V}_\mu \ket{\rm HF}$. 
CSFs are eigenstates of the total spin $\hat{S}^2$.

\begin{figure*}
    \centering
    \includegraphics[width=0.9\textwidth]{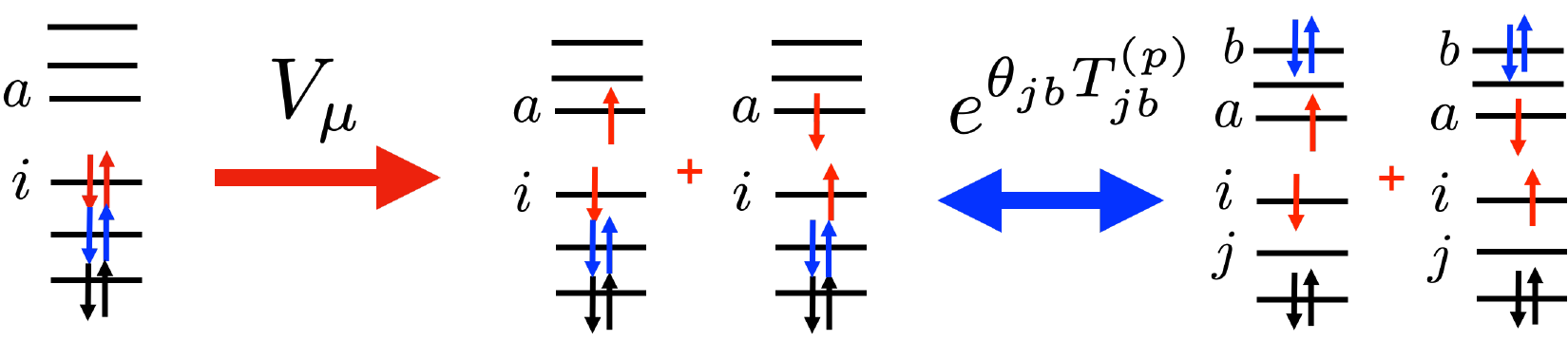}
    \caption{Illustration of a singlet CSF with two unpaired electrons ($\Omega=2$) and pair rotations.}
    \label{fig:CSFW}
\end{figure*}

The Q-SENSE formalism can be applied to states of arbitrary spin and seniorities. Here, for concreteness, we focus on singlet CSFs with seniority up to 4:
\bea \notag
\{\ket{\rm CSF_\mu}\} = \{\ket{\text{HF}}&,&\hat{E}_{ia}^{0,0}\ket{\text{HF}},
\hat{E}_{jb}^{0,0}\hat{E}_{ia}^{0,0}\ket{\text{HF}}, \\ 
\frac{1}{\sqrt{3}}\Big(-\hat{E}_{jb}^{1,1}\hat{E}_{ia}^{1,-1}&+&\hat{E}_{jb}^{1,0}\hat{E}_{ia}^{1,0}- \hat{E}_{jb}^{1,-1}\hat{E}_{ia}^{1,1}\Big)\ket{\text{HF}}\},
\label{eq:CSFs}
\eea
where spherical tensor excitation operators are used
\bea \notag
    \hat{E}_{ia}^{0,0} &=& \frac{1}{\sqrt{2}}\big(\hat{a}_{a\downarrow}^\dagger \hat{a}_{i\downarrow} + \hat{a}_{a\uparrow}^\dagger \hat{a}_{i\uparrow} \big) \\ \notag
    \hat{E}_{ia}^{1,0} &=& \frac{1}{\sqrt{2}}\big(\hat{a}_{a\downarrow}^\dagger \hat{a}_{i\downarrow} - \hat{a}_{a\uparrow}^\dagger \hat{a}_{i\uparrow}\big) \\
    \hat{E}_{ia}^{1,1} &=& -\hat{a}_{a \uparrow}^\dagger \hat{a}_{i \downarrow}, \quad \hat{E}_{ia}^{1,-1} = \hat{a}_{a \downarrow}^\dagger \hat{a}_{i \uparrow}. \label{eq:exc}
\eea
Figure~\ref{fig:CSFW} illustrates a CSF with two unpaired electrons and a pair rotation.  
In principle, one can go further by unpairing more electrons in singlet CSFs; this would require additional combinations of spherical tensor operators to preserve $\hat{S}^2$. The choice of initial CSFs (i.e., the $\{\hat V_\mu\}$ set) is one of the degrees of freedom in Q-SENSE. Note that the set of singly occupied orbitals is fully determined by $\hat{V}_\mu$.

The Q-SENSE basis states $\left\{\ket{\phi_\mu}\right\}$ are eigenstates of a set of mutually commuting orbital seniority operators,
\begin{equation}
    \hat{\Omega}_i = (\hat{n}_{i\uparrow} + \hat{n}_{i\downarrow})(2 - \hat{n}_{i\uparrow} - \hat{n}_{i\downarrow}), \label{orbital_seniority}
\end{equation}
where $\hat{n}_{i\sigma} = \hat{a}_{i\sigma}^\dagger \hat{a}_{i\sigma}$. The operator $\hat{\Omega}_i$ counts the number of unpaired electrons on the $i^{\rm th}$ molecular orbital. By trivial extension, $\left\{\ket{\phi_\mu}\right\}$ are also eigenstates of the total seniority operator
\begin{equation}
    \hat{\Omega} = \sum_{i=1}^{N_\text{orb}} \hat{\Omega}_i,
\end{equation}
where $N_\text{orb}$ denotes the number of molecular orbitals (MOs).

The space spanned by $\left\{\ket{\phi_\mu}\right\}$ can be extended to the complete singlet Hilbert subspace of $N_e$ electrons if: 1) repetitions 
of pair rotations in $\hat{W}_{\mu}$ are allowed; 2) singlet CSFs built by $\hat{V}_\mu$ span all possible combinations and singlet coupling of unpaired electrons 
on all available orbitals.
The first condition holds because $\hat{T}_{rs}^{(p)}$ forms a universal set for the zero-seniority subspace, which is equivalent to the Doubly Occupied Configuration Interaction (DOCI) space \cite{Henderson2014DOCI}. The second condition stems from the inability of $\hat{W}_\mu$ to move unpaired electrons formed by $\hat{V}_\mu$. This full extension will require an exponential number of parameters; thus, in applications, subsets of states $\ket{\phi_\mu}$ are selected using heuristic choices that are presented in Section~\ref{sec:basis_selection}.

\subsection{Matrix Element Estimation in Q-SENSE}\label{sec:estimation_general}

The construction of the Q-SENSE basis set $\left\{\ket{\phi_\mu}\right\}$ allows for substantial savings in estimating Hamiltonian matrix elements. The key point is that certain aspects of the matrix element estimation can be treated efficiently on a classical computer, thereby reducing the cost of the subsequent quantum computation. These reductions are similar in spirit to qubit tapering \cite{bravyiTaperingQubitsSimulate2017}.

To simulate electronic structure on a quantum computer, the molecular electronic Hamiltonian is mapped to a qubit Hamiltonian via a fermion-to-qubit mapping such as the Jordan--Wigner transformation \cite{jordanUeberPaulischeAequivalenzverbot1928}. In qubit form, the electronic Hamiltonian is expressed as a linear combination of Pauli products on $2N_\text{orb}$ qubits. We adopt the convention that the $\uparrow$ and $\downarrow$ spin-orbitals of the $i^{\text{th}}$ molecular orbital are mapped to qubits $2i$ and $2i+1$, respectively.

Formally, the need to use a quantum computer in Q-SENSE stems from the unitary $\hat{W}_\mu$ used to construct the basis states, as it cannot be implemented efficiently on a classical computer. However, $\hat{W}_\mu$ is more structured than a general number-conserving unitary, such as one constructed via a UCC ansatz, in two ways. First, $\hat{W}_\mu$ is built exclusively from singlet pair-excitation operators $\hat{T}_{rs}^{(p)}$. Second, in general, $\hat{W}_\mu$ acts only on a restricted subset of all MOs. Let $\mathcal{S}_W$, $\mathcal{S}_V$, and $\mathcal{S}_N$ denote the MOs on which $\hat{W}_\mu$ acts, $\hat{V}_\mu$ acts, and neither acts, respectively. By construction, these three sets are disjoint. For instance, for the singlet CSF described in Figure \ref{fig:CSFW}, $\mathcal S_V = \{i, a\}$ and $\mathcal S_W = \{j, b\}$ correspond to the sets of MOs on which $\hat V_\mu$ and $\hat W_\mu$ act, respectively, while $\mathcal S_N$ contains the remaining MOs that are unaffected by $\hat W_\mu \hat V_\mu$.

The disjointness of the MO subsets implies that each Q-SENSE basis state factorizes into parts that can be implemented efficiently on a classical computer and a single part encoding the action of $\hat{W}_\mu$ that requires a quantum computer. To obtain this factorization, we apply the Jordan--Wigner transformation, which maps the orbital-seniority operator to
\begin{equation}
    \hat{\Omega}_i = \frac{\hat{1} - \hat{z}_{2i}\hat{z}_{2i+1}}{2}.
\end{equation}
Importantly, it is mapped to a single Pauli operator, up to constants. This fact allows us to halve the number of qubits needed to encode the Q-SENSE states using a Clifford unitary $\hat{U}_{\rm c}$ that satisfies
\begin{equation}
    \hat{U}_{\rm c} \hat{z}_{2i}\hat{z}_{2i+1} \hat{U}_{\rm c}^\dagger = \hat{z}_{N_\text{orb} + i}.
\end{equation}
Therefore, for a Q-SENSE basis state $\ket{\phi_\mu}$ such that $\hat{\Omega}_i \ket{\phi_\mu} = v_i\ket{\phi_\mu}$, we have
\begin{equation}
    \hat{U}_{\rm c} \ket{\phi_\mu} =  \ket{\phi_\mu^{(c)}}\ket{\vec{v}}.
\end{equation}
The orbital seniorities $\vec{v}$ are stored in the computational basis $\ket{\vec{v}}$, which can be treated classically, while $\ket{\phi_\mu^{(c)}}$ is an $N_\text{orb}$-qubit state that encodes $\ket{\phi_\mu}$ up to its seniority values. 

Each qubit in the compressed Q-SENSE basis state corresponds to an MO. Given the disjointness of $\mathcal{S}_W, \mathcal{S}_V, \mathcal{S}_N$, it follows that $\ket{\phi_\mu^{(c)}}$ factorizes over the corresponding subsets:
\begin{equation}
    \ket{\phi_\mu^{(c)}} = \ket{\phi_\mu^{(\mathcal{S}_W)}}\ket{\phi_\mu^{(\mathcal{S}_V)}}\ket{\phi_\mu^{(\mathcal{S}_N)}}.
\end{equation}
Here, $\ket{\phi_\mu^{(\mathcal{S}_W)}}$ is an $|\mathcal{S}_W|$-qubit state and is the only classically intractable factor. The state $\ket{\phi_\mu^{(\mathcal{S}_V)}}$ corresponds to the CSF implemented by $\hat{V}_\mu$ in the Clifford-transformed basis, while $\ket{\phi_\mu^{(\mathcal{S}_N)}}$ is a computational basis state encoding the Hartree--Fock occupations on MOs in the set $\mathcal{S}_N$. Together with the seniority-encoding state $\ket{\vec{v}}$, these latter components are classically tractable and can be prepared efficiently on a quantum computer.

The factorization of each Q-SENSE basis state into classically tractable and intractable parts motivates a hybrid quantum-classical approach to estimating the matrix element between two such states, $\ket{\phi_\mu}$ and $\ket{\phi_\nu}$. The first step is to partition the qubits into two subsets: (1) a subset $\mathcal{C}$ that collects those qubits whose contributions to $\braket{\phi_\mu|\hat{H}|\phi_\nu}$ can be evaluated efficiently on a classical computer; and (2) the complementary subset $\mathcal{Q}$, which requires a quantum computer. Formally, these subsets must satisfy two conditions. First, both Q-SENSE states factorize over $\mathcal{C}$ and $\mathcal{Q}$:
\begin{align}
    \hat{U}_{\rm c}\ket{\phi_\mu} &= \ket{\phi_\mu^{(\mathcal{C})}}\ket{\phi_\mu^{(\mathcal{Q})}},\\
    \hat{U}_{\rm c}\ket{\phi_\nu} &= \ket{\phi_\nu^{(\mathcal{C})}}\ket{\phi_\nu^{(\mathcal{Q})}}.
\end{align}
Second, for any Pauli operator $\hat{P}$ that acts on qubits in $\mathcal{C}$, the matrix element $\braket{\phi_\mu^{(\mathcal{C})}|\hat{P}|\phi_\nu^{(\mathcal{C})}}$ can be evaluated efficiently on a classical computer. The procedure for obtaining the $\{\mathcal{C},\mathcal{Q}\}$ partition from the underlying $\{\mathcal{S}_W,\mathcal{S}_V,\mathcal{S}_N\}$ partition of both states is described in Section \ref{sec:cq_part}. Using the $\{\mathcal{C},\mathcal{Q}\}$ partition, we can express $\braket{\phi_\mu|\hat{H}|\phi_\nu}$ as a matrix element of an effective Hamiltonian on $|\mathcal{Q}|$ qubits. This is done by first writing the Hamiltonian in the Clifford-transformed basis used to factorize the seniority-encoding state,
\begin{equation}
    \hat{U}_c\hat{H}\hat{U}_c^\dagger = \sum_k c_k \hat{P}_k,
\end{equation}
and then noting that each Pauli term $\hat{P}_k$ also factorizes over the partition:
\begin{equation}
    \hat{P}_k = \hat{P}_k^{(\mathcal{C})} \otimes \hat{P}_k^{(\mathcal{Q})}.
\end{equation}
Therefore,
\begin{align}
    \braket{\phi_\mu|\hat{H}|\phi_\nu} &= \braket{\phi_\mu^{(\mathcal{Q})}|\hat{h}_{\mu\nu}|\phi_\nu^{(\mathcal{Q})}},\\
    \hat{h}_{\mu\nu} &= \sum_k c_k \braket{\phi_\mu^{(\mathcal{C})}|\hat{P}_k^{(\mathcal{C})}|\phi_\nu^{(\mathcal{C})}} \hat{P}_k^{(\mathcal{Q})}.
\end{align}
The effective Hamiltonian $\hat{h}_{\mu\nu}$ obtained by contracting qubits in $\mathcal{C}$ is much simpler than the original Hamiltonian, leading to substantial savings in measurement cost. The first reason is a reduction in the number of qubits required to encode the effective Hamiltonian and state. In the worst case, $|\mathcal{Q}| = |\mathcal{C}| = N_\text{orb}$, since the qubits that store the seniorities can always be treated classically. The second reason is that expectation values of $\hat{h}_{\mu\nu}$ are much simpler to estimate. This follows because the procedure effectively removes Pauli terms from $\hat{H}$ for which $\braket{\phi_\mu^{(\mathcal{C})}|\hat{P}_k^{(\mathcal{C})}|\phi_\nu^{(\mathcal{C})}} = 0$, and combines distinct Pauli terms $\hat{P}_k$ and $\hat{P}_l$ from $\hat{H}$ into a single Pauli term in $\hat{h}_{\mu\nu}$ whenever $\hat{P}_k^{(\mathcal{Q})} = \hat{P}_l^{(\mathcal{Q})}$. Since any single Pauli product maps a computational basis state $\ket{\vec{z}}$ to a single basis state $\ket{\vec{z}'} \propto \hat{P}_k^{(\mathcal{C})}\ket{\vec{z}}$ among $2^{|\mathcal{C}|}$ possibilities, and the classical states $\ket{\phi_\mu^{(\mathcal{C})}}$ and $\ket{\phi_\nu^{(\mathcal{C})}}$ are themselves very sparse, the matrix element $\braket{\phi_\mu^{(\mathcal{C})}|\hat{P}_k^{(\mathcal{C})}|\phi_\nu^{(\mathcal{C})}}$ is exponentially unlikely to be nonzero. Consequently, the effective Hamiltonian is much sparser than the original electronic Hamiltonian, and therefore easier to measure. Note also that, in general, the effective Hamiltonian may be non-Hermitian for off-diagonal matrix elements, since $\braket{\phi_\mu^{(\mathcal{C})}|\hat{P}_k^{(\mathcal{C})}|\phi_\nu^{(\mathcal{C})}}$ can be imaginary when $\hat{P}_k^{(\mathcal{C})}$ contains an odd number of Pauli $\hat{y}$ factors.

\section{Technical Details of the Q-SENSE Algorithm}\label{sec:technical}

In this section, we present details of various technical aspects of the Q-SENSE algorithm. 

\subsection{Obtaining the Q-SENSE Basis States}\label{sec:basis_selection}

Each ansatz state involves two unitary transformations, $\hat V_\mu$ and $\hat W_\mu$. Here we detail how to choose their parameters.
Choosing $\hat V_\mu$ involves two decisions: (1) which orbitals host the unpaired electrons (the orbital window), and (2) how many unpaired electrons are allowed (the seniority level). Every $S^2$ eigenstate with a fixed number of unpaired electrons has a finite number of combinations, for example, 
in the singlet case, two unpaired electrons have only one singlet while four unpaired electrons have two singlet configurations. 
The parameters of $\hat W_\mu$ depend on $\hat V_\mu$, but in all cases two questions arise: (1) which pairs $(r,s)$ to choose for the generators $\hat T_{rs}^{(p)}$ in $\hat W_\mu$ [\eq{eq:W}], and (2) how to determine the corresponding amplitudes.

Choices in the definition of $\hat V_\mu$ will determine the number of Hamiltonian matrix elements as well as the quantum circuit sizes (mainly due to higher circuit cost to prepare higher seniority CSFs; see Section \ref{sec:circuits} for details). The definition of $\hat W_\mu$ also affects circuit cost, which grows with the number of generators $\hat T_{rs}^{(p)}$. If circuits for $\hat W_\mu$ exceed current NISQ capabilities, the circuit depth can be reduced by promoting generators in $\hat{W}_\mu$ to excitations used to produce additional CSFs with unchanged seniority eigenvalues, e.g., $\ket{\text{CSF}_\mu^{(ai)}} = \hat T_{ai}^{(p)}\hat V_\mu\ket{\rm HF}$.

\begin{figure}
    \centering
    \includegraphics[width=0.8\columnwidth]{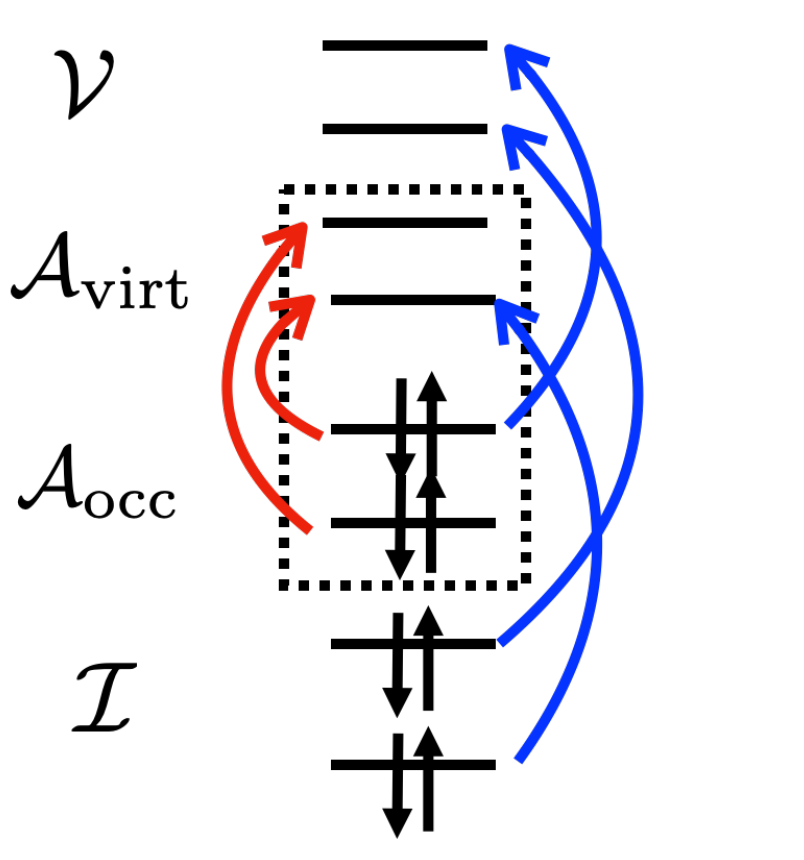}
    \caption{Orbital partitioning to virtual orbitals ($\mathcal{V}$), active orbitals (dashed box, $\mathcal{A} = \mathcal{A}_\text{occ} \cup \mathcal{A}_\text{virt}$), inactive orbitals ($\mathcal{I}$); blue and red arrows correspond to external and internal excitations.}
    \label{fig:orbs}
\end{figure}

We propose a heuristic selection of $\hat V_\mu$ based on the Hartree-Fock orbital energies, which allows us to define an active orbital set $\mathcal{A}$ near the Fermi level.   
The orbitals lying below $\mathcal{A}$ form the set of inactive orbitals $\mathcal{I}$, and the orbitals lying above $\mathcal{A}$ form the set of virtual orbitals $\mathcal{V}$ (see \fig{fig:orbs}). 
The active space itself admits a decomposition $\mathcal{A} = \mathcal{A}_\text{occ} \cup \mathcal{A}_\text{virt}$, where $\mathcal{A}_\text{occ}$ 
denotes active space orbitals below the Fermi level, which are doubly occupied in $\ket{\rm HF}$, and $\mathcal{A}_\text{virt}$ denotes active space orbitals above the 
Fermi level, which are empty in $\ket{\rm HF}$. We call excitations in \eq{eq:exc} ``internal excitations'' 
if $i,a \in \mathcal{A}$, and ``external excitations'' otherwise. 

To generate an initial set of CSFs ${\hat{V}_\mu\ket{\mathrm{HF}}}$, we first construct $N_\text{ini}$ CSFs as in Eq.~(\ref{eq:CSFs}) using internal excitations of $\ket{\mathrm{HF}}$ with $i,j\in\mathcal{A}_\text{occ}$ and $a,b\in\mathcal{A}_\text{virt}$. Second, we solve the eigenvalue problem in the subspace spanned by these CSFs and retain only those with non-negligible ground-state weight, i.e., $|c_\mu|^2 > \epsilon_1$.

For each $\ket{\rm CSF_\mu}$, we also construct an extension set $\mathcal{E}_\mu$ containing additional CSFs which significantly contribute to the ground state. To do this, we consider all $(a,i)$ pairs, and obtain the pair-excited $\ket{\text{CSF}_{\mu}^{(ai)}}=\hat T_{ai}^{(p)}\ket{\rm CSF_\mu}$ state, which together with $\{\ket{\rm CSF_\mu}\}_{\mu=1}^{N_\text{ini}}$ forms an $(N_\text{ini}+1)$-dimensional subspace. We evaluate the ground-state energy difference between the $(N_\text{ini}+1)$ and $N_\text{ini}$-dimensional subspaces, denoted $\Delta E_{ai,\mu}$. If $|\Delta E_{ai,\mu}| > \epsilon_2$, the $(a,i)$ pair is included in $\mathcal{E}_{\mu}$.  

We explore two strategies for obtaining the generators and angles of $\hat{W}_\mu$, which differ in how they exploit the extension sets $\mathcal{E}_\mu$: (1) variational optimization (VO), and (2) perturbation theory (PT).

\textbf{Variational Optimization:} The $(a,i)$ pairs from $\mathcal{E}_\mu$ are used to construct $\hat W_\mu$ [\eq{eq:W}], such that the corresponding generators are ordered by decreasing $|\Delta E_{ai,\mu}|$, so that pairs with larger $|\Delta E_{ai,\mu}|$ appear closer to $\ket{\text{CSF}_\mu}$ in the product. The initial rotation amplitudes $\left\{\theta_{ai}\right\}$ are variationally optimized to minimize the lowest eigenvalue of \eq{GEVP}. 

\textbf{Perturbation Theory:} The PT approach yields a reduction in circuit depth and optimization costs compared to VO by using a larger initial CSF set and shorter $\hat W_\mu$ circuits. The $\{\ket{\rm CSF_\mu}\}_{\mu=1}^{N_\text{ini}}$ set is enlarged by including $\hat T_{ai}^{(p)}\ket{\rm CSF_\mu}$ for {\it all internal} $(a,i)$ excitation pairs in $\mathcal{E}_{\mu}$. This produces an extended set $\{\ket{\text{CSF}_\mu^{(ai)}}\}$ of initial CSFs. The pair-rotations $\hat W_\mu$ are built using {\it only external} $(b,j)$ pairs in $\mathcal{E}_{\mu}$. The corresponding pair-rotation amplitudes are set to the ground-state MP2 amplitudes. $\hat W_\mu$ is used for all $\ket{\text{CSF}_\mu^{(ai)}}$ to generate the final subspace for \eq{eq:qse}, $\ket{\phi_\nu} = \hat W_\mu \ket{\text{CSF}_\mu^{(ai)}}$, where the subscript $\nu$ is a composite index of $\mu$ and $(ai)$.

Lastly, in addition to $\hat W_\mu$ and $\hat V_\mu$ we introduce a common orbital rotation unitary in both VO and PT methods
\begin{equation}
\hat U_{\rm orb}(\bar{t}) = \prod_{pq,\sigma} e^{t_{pq}(\hat a_{p\sigma}^\dagger \hat a_{q\sigma} - h.c.)}
\end{equation}
to lower the energy even further,
\begin{equation}
E_{\min} = \min_{\bar{t},\bar{c}} \sum_{\mu,\nu} c_\mu^* c_\nu \bra{\phi_\mu} \hat U_{\rm orb}(\bar{t})^\dagger \hat H \hat U_{\rm orb}(\bar{t}) \ket{\phi_\nu}. \label{eq:orbital_opt}
\end{equation}
Implementing $\hat U_{\rm orb}(\bar{t})$ does not require additional quantum circuits, since the Hamiltonian transformation 
$\tilde H(\bar{t}) =  \hat U_{\rm orb}(\bar{t})^\dagger \hat H \hat U_{\rm orb}(\bar{t})$ can be performed efficiently on a classical computer and does not change the Hamiltonian's form.  
However, minimizing the energy with respect to the orbital-rotation amplitudes $\{t_{pq}\}$ requires solving an eigenvalue problem for the transformed Hamiltonians $\tilde H(\bar{t})$. This relaxation reduces the number of matrix elements and circuit depths needed to reach a target accuracy, but increases the number of quantum measurements because Hamiltonian matrix elements must be constructed repeatedly.

\subsection{Details of Matrix Element Estimation}\label{sec:estimation}

\subsubsection{Obtaining Classical/Quantum Partitioning}\label{sec:cq_part}

Here, we explain how to partition the qubits into $\mathcal{C}$ and $\mathcal{Q}$ when evaluating the matrix element between two Q-SENSE basis states $\ket{\phi_\mu}$ and $\ket{\phi_\nu}$. Our goal is to maximize $|\mathcal{C}|$, thereby minimizing the quantum resources required for each matrix element. In what follows, we use the notation $\ket{\phi_\tau^{(\mathcal{X})}}$, for $\tau \in \{\mu, \nu\}$, to denote the factor of $\ket{\phi_\tau}$ on qubits $\mathcal{X}$. In the Clifford-transformed basis that maps orbital seniorities to single-qubit $\hat{z}$ operators, each Q-SENSE basis state factorizes as
\begin{align}
    \hat{U}_{\rm c}\ket{\phi_\mu} &= \ket{\phi_\mu^{(\mathcal{S}_W)}}\ket{\phi_\mu^{(\mathcal{S}_V)}}\ket{\phi_\mu^{(\mathcal{S}_N)}}\ket{\vec{v}},\nonumber\\
    \hat{U}_{\rm c}\ket{\phi_\nu} &= \ket{\phi_\nu^{(\mathcal{S}_W')}}\ket{\phi_\nu^{(\mathcal{S}_V')}}\ket{\phi_\nu^{(\mathcal{S}_N')}}\ket{\vec{w}},\label{eq:first_factorization}
\end{align}
where $\ket{\vec{v}}$ and $\ket{\vec{w}}$ encode the orbital seniorities. In general, Q-SENSE basis states admit more factors than the four shown in Eq.~\eqref{eq:first_factorization}. For example, for $\ket{\phi_\mu}$, both $\ket{\vec{v}}$ and $\ket{\phi_\mu^{(\mathcal{S}_N)}}$ are computational basis states and are therefore further factorizable into tensor products of single-qubit $\ket{0}$ and $\ket{1}$ states. Similarly, depending on the form of $\hat{V}_{\mu}$, it may be that $\ket{\phi_\mu^{(\mathcal{S}_V)}}$ further factorizes into tensor products of simpler states. The classically hard factor $\ket{\phi_\mu^{(\mathcal{S}_W)}}$ may also admit additional factorizations, depending on the generators in $\hat{W}_\mu$, but it is not necessary to consider these in the subsequent derivation. Therefore, we assume that the classically hard factors are not further factorizable.

To proceed, we decompose each Q-SENSE basis state separately into a maximal tensor product of factors that cannot be further factorized without additional unitary transformations:
\begin{align}
    \hat{U}_{\rm c}\ket{\phi_\mu} &= \bigotimes_{k=1}^{K} \ket{\phi_\mu^{(\mathcal{C}_k)}}\ket{\phi_\mu^{(\mathcal{S}_W)}},\nonumber\\
    \hat{U}_{\rm c}\ket{\phi_\nu} &= \bigotimes_{l=1}^{L} \ket{\phi_\nu^{(\mathcal{C}_l')}}\ket{\phi_\nu^{(\mathcal{S}_W')}}.
\end{align}
Here, $\mathcal{C}_k$ and $\mathcal{C}_l'$ denote the resulting minimal subsets of qubits over which the Q-SENSE basis states factorize. Since $\ket{\phi_\mu}$ and $\ket{\phi_\nu}$ must both factorize over $\mathcal{C}$ and $\mathcal{Q}$, we now form a common partition $\{\mathcal{P}_m\}_{m=1}^{M}$ of all qubits that is compatible with both factorizations. Concretely, $\mathcal{P}_m$ is taken as the finest common coarsening (i.e., the “join”) of $\{\mathcal{S}_W\} \cup \{\mathcal{C}_k\}_{k=1}^{K}$ and $\{\mathcal{S}_W'\} \cup \{\mathcal{C}_l'\}_{l=1}^{L}$. The partition $\{\mathcal{P}_m\}_{m=1}^{M}$ can be characterized by two properties: (1) every $\mathcal{P}_m$ is a union of elements in $\{\mathcal{S}_W\} \cup \{\mathcal{C}_k\}_{k=1}^{K}$ that can also be written as a union of elements in $\{\mathcal{S}_W'\} \cup \{\mathcal{C}_l'\}_{l=1}^{L}$, and (2) the size $M$ of $\{\mathcal{P}_m\}_{m=1}^{M}$ is maximal among all partitions satisfying condition 1. Condition 2 ensures that the sets $\mathcal{P}_m$ are minimal with respect to this property. By construction, we now have a unified factorization of both Q-SENSE basis states:
\begin{align}
    \hat{U}_{\rm c}\ket{\phi_\mu} &= \bigotimes_{m=1}^{M} \ket{\phi_\mu^{(\mathcal{P}_m)}},\nonumber\\
    \hat{U}_{\rm c}\ket{\phi_\nu} &= \bigotimes_{m=1}^{M} \ket{\phi_\nu^{(\mathcal{P}_m)}}.
\end{align}

We now define which qubits can be treated classically and which must be treated on the quantum computer: qubits in $\mathcal{P}_m$ are treated on the quantum computer if and only if $\mathcal{S}_W \subset \mathcal{P}_m$ or $\mathcal{S}_W' \subset \mathcal{P}_m$. In the case that $\mathcal{S}_W \cap \mathcal{S}_W' \neq \varnothing$, which is common in our construction of Q-SENSE basis states, it follows that $\mathcal{S}_W \cup \mathcal{S}_W'$ is contained in a single $\mathcal{P}_m$; without loss of generality, we label this set as $\mathcal{P}_1$. Then $\mathcal{Q} = \mathcal{P}_1$ and $\mathcal{C} = \bigcup_{m=2}^{M}\mathcal{P}_m$. In the case that $\mathcal{S}_W \cap \mathcal{S}_W' = \varnothing$, it is possible (but not required) that $\mathcal{S}_W$ and $\mathcal{S}_W'$ are contained in separate $\mathcal{P}_m$; without loss of generality, we label these two sets as $\mathcal{P}_1$ and $\mathcal{P}_2$. Then $\mathcal{Q} = \mathcal{P}_1 \cup \mathcal{P}_2$ and $\mathcal{C} = \bigcup_{m=3}^{M} \mathcal{P}_m$.

\subsubsection{Extended Swap Test Formalism}\label{sec:swap_test}

To estimate off-diagonal elements, we use the extended swap-test formalism of Ref.~\cite{parrishQuantumFilterDiagonalization2019}, which employs a single ancilla qubit to map an off-diagonal matrix element to an expectation value:
\begin{equation}
    \braket{\phi_\mu^{(\mathcal{Q})}|\hat{h}_{\mu\nu}|\phi_\nu^{(\mathcal{Q})}} = \braket{\Phi_{\mu\nu}^{(\mathcal{Q})}|\hat{h}'_{\mu\nu}|\Phi_{\mu\nu}^{(\mathcal{Q})}}. \label{swap_test_general}
\end{equation}
Here,
\begin{equation}
    \ket{\Phi_{\mu\nu}^{(\mathcal{Q})}} := \frac{1}{\sqrt{2}}\left(\ket{0}\ket{\phi_\mu^{(\mathcal{Q})}} + \ket{1}\ket{\phi_\nu^{(\mathcal{Q})}}\right). \label{swap_test_state}
\end{equation}
A valid choice for $\hat{h}'_{\mu\nu}$, without making assumptions about $\hat{h}_{\mu\nu}$, $\ket{\phi_\mu^{(\mathcal{Q})}}$, or $\ket{\phi_\nu^{(\mathcal{Q})}}$, is $\hat{h}'_{\mu\nu} = (\hat{x} + \imag \hat{y}) \otimes \hat{h}_{\mu\nu}$, which has twice as many Pauli terms as $\hat{h}_{\mu\nu}$. In practice, for each Pauli term $\hat{P}$ in $\hat{h}_{\mu\nu}$, only the $\hat{x}$ contribution is needed when $\braket{\phi_\mu^{(\mathcal{Q})}|\hat{P}|\phi_\nu^{(\mathcal{Q})}}$ is real, and only the $\imag \hat{y}$ contribution is needed when $\braket{\phi_\mu^{(\mathcal{Q})}|\hat{P}|\phi_\nu^{(\mathcal{Q})}}$ is imaginary. Whether the matrix element is real or imaginary can be determined classically by checking the parity of the number of Pauli $\hat{y}$ operators in $\hat{P}$. Therefore, our construction of $\hat{h}'_{\mu\nu}$ requires only a single additional Pauli product compared to $\hat{h}_{\mu\nu}$, since the constant term in $\hat{h}_{\mu\nu}$ is mapped to the Pauli operator $\hat{x} \otimes \hat{1}^{|\mathcal{Q}|}$ in $\hat{h}_{\mu\nu}'$.

\subsubsection{Constant Term Optimization}\label{sec:constant}

In any quantum subspace expansion method that uses orthogonal states, such as Q-SENSE, one can treat the constant term as a degree of freedom for optimizing the sampling cost of off-diagonal matrix elements. We describe this optimization here before addressing technical aspects of its application to Q-SENSE.

Consider two $N$-qubit orthogonal states $\ket{\psi}$ and $\ket{\phi}$, for which we wish to obtain the matrix element $T = \braket{\psi|\hat{O} + C\hat{1}^N|\phi}$. Here, we separate out the constant term $C\hat{1}^N$, so that $\hat{O}$ is a qubit Hamiltonian without a constant term. For simplicity, assume that $\ket{\psi}$, $\ket{\phi}$, $\hat{O}$, and $C$ are all real-valued. Defining $\ket{\Psi} = 2^{-1/2}(\ket{0}\ket{\psi} + \ket{1}\ket{\phi})$, the extended swap-test formalism expresses the matrix element in terms of the following expectation value:
\begin{equation}
    T = \braket{\Psi|\hat{x} \otimes \hat{O} + C\hat{x}\otimes \hat{1}^N|\Psi}.
\end{equation}
Within the extended swap-test formalism, the orthogonality relation $\braket{\psi|\phi} = 0$ is encoded in the expectation value
\begin{equation}
    \braket{\Psi|\hat{x}\otimes\hat{1}^N|\Psi} = 0.
\end{equation}
Therefore, the Pauli operator $\hat{x} \otimes \hat{1}^{N}$ has mean zero but nonzero variance, and consequently nonzero covariance with the rest of the Hamiltonian. The variance $V(C)$ of the matrix-element $T$ estimator as a function of the constant $C$ is given by
\begin{align}
    V(C) &= C^2 + C\Big(\braket{\psi|\hat{O}|\psi} + \braket{\phi|\hat{O}|\phi}\Big)\nonumber\\
         &\quad + \frac{1}{2}\Big(\braket{\psi|\hat{O}^2|\psi} + \braket{\phi|\hat{O}^2|\phi}\Big) - T^2.
\end{align}
We can then adjust the coefficient $C$ to minimize the variance of the estimator for $T$ without affecting its mean value. The optimal value is
\begin{equation}
    C^{(\text{min})} = -\frac{1}{2} \big(\braket{\psi|\hat{O}|\psi} + \braket{\phi|\hat{O}|\phi}\big). \label{constant_shift_opt}
\end{equation}
This optimization requires knowledge of the diagonal elements of the subspace Hamiltonian, which can be obtained by estimating them before the off-diagonal elements.

Orthogonality of two Q-SENSE basis states $\ket{\phi_\mu}$ and $\ket{\phi_\nu}$ implies orthogonality of either their factor on qubits $\mathcal{C}$ or their factor on qubits $\mathcal{Q}$:
\begin{equation}
    \braket{\phi_\mu|\phi_\nu} = \braket{\phi_\mu^{(\mathcal{C})}|\phi_\nu^{(\mathcal{C})}}\braket{\phi_\mu^{(\mathcal{Q})}|\phi_\nu^{(\mathcal{Q})}} = 0.
\end{equation}
When $\braket{\phi_\mu^{(\mathcal{C})}|\phi_\nu^{(\mathcal{C})}} = 0$, the constant term of the Hamiltonian is removed via the classical partial matrix-element evaluation. Otherwise, if $\braket{\phi_\mu^{(\mathcal{C})}|\phi_\nu^{(\mathcal{C})}} \neq 0$, the constant term survives the classical partial matrix-element evaluation and contributes nontrivially to the resulting effective Hamiltonian $\hat{h}_{\mu\nu}$. In this case, $\braket{\phi_\mu^{(\mathcal{Q})}|\phi_\nu^{(\mathcal{Q})}} = 0$, and the constant-term optimization is applicable.

\subsubsection{Hamiltonian Partitioning into Measurable Fragments}\label{sec:partitioning}

The expectation values $\braket{\Phi_{\mu\nu}^{(\mathcal{Q})}|\hat{h}'_{\mu\nu}|\Phi_{\mu\nu}^{(\mathcal{Q})}}$ are estimated via quantum measurements. Since quantum computers natively measure in the Pauli $\hat{z}$ basis, we estimate the desired matrix elements by first partitioning $\hat{h}_{\mu\nu}'$ into a sum of diagonalizable fragments,
\begin{equation}
    \hat{h}'_{\mu\nu} = \sum_\alpha \hat{F}_{\mu\nu}^{(\alpha)}. \label{partitioning}
\end{equation}
The expectation value of $\hat{h}_{\mu\nu}'$ is then obtained as the sum of the expectation values of the fragments $\hat{F}_{\mu\nu}^{(\alpha)}$. We use fully-commuting (FC) fragments, each constructed as a linear combination of mutually commuting Pauli products, and therefore mappable to a measurable Ising form by applying a Clifford transformation \cite{yenMeasuringAllCompatible2020}. To obtain the fragments, we use the sorted-insertion (SI) algorithm, which has been shown to yield decompositions with lower measurement cost than those produced by comparable decomposition algorithms \cite{crawfordEfficientQuantumMeasurement2021}. In SI, the Hamiltonian’s Pauli terms are first sorted in descending order of their coefficient magnitudes. The algorithm then iterates through this sorted list, adding each term to the first existing fragment with which it fully commutes; if a term does not commute with any existing fragment, it is used to initiate a new fragment.

\subsection{Quantum Circuits for Estimating Q-SENSE Matrix Elements}\label{sec:circuits}

In this section, we describe the quantum circuits used to prepare the compressed Q-SENSE basis states $\ket{\phi_\mu^{(\mathcal Q)}}$ and $\ket{\Phi_{\mu\nu}^{(\mathcal Q)}}$, which are obtained by applying the Clifford transformation $\hat{U}_c$ and removing classically evaluable components. As described in Sec.~\ref{sec:cq_part},
$\ket{\phi_\mu^{(\mathcal Q)}}$ can be a tensor product of separable parts of $\ket{\phi_\mu^{(\mathcal S_V)}}$, $\ket{\phi_\mu^{(\mathcal S_W)}}$, and $\ket{\phi_\mu^{(\mathcal S_N)}}$, and is therefore constructed as a product of unitaries
\begin{equation}
    \ket{\phi_\mu^{(\mathcal Q)}} = \hat W_\mu^{(c)} \hat S_\mu \hat D_\mu \ket{0}^{\otimes |\mathcal Q|}.
\end{equation}
Here, $\hat S_\mu$ prepares $\ket{\phi_\mu^{(\mathcal S_V)}}$, $\hat D_\mu$ prepares the occupations of unoccupied and doubly occupied MOs prior to the action of electron-pair excitations, and $\hat W_\mu^{(c)}$ is a sequence of compressed electron-pair excitations.

Due to seniority restrictions, $\ket{\phi_\mu^{(\mathcal S_V)}}$ takes one of four distinct forms, whose operator definitions are provided in \eq{eq:CSFs}. We use $\hat S_{\mathcal S_V}^{(0-3)}$ to label the corresponding four circuits. The quantum circuits for $\hat S_{\mathcal S_V}^{(1-3)}$ are shown in Fig.~\ref{fig:csf_circuits}. States with seniority $\Omega = 0$ contain no singly occupied orbitals and are prepared by $\hat S^{(0)} = \hat 1$. States with two unpaired electrons ($\Omega = 2$) are generated by applying the operator $\hat E_{ia}^{0,0}$ to the Hartree--Fock state and are prepared using $\hat S_{ia}^{(1)}$ applied to qubits $\mathcal S_V = \{i, a\}$ initialized to $\ket{0}$. There are two types of states with four unpaired electrons ($\Omega = 4$). A CSF of the first type is generated by applying the operator $\hat E_{ia}^{0,0}\hat E_{jb}^{0,0}$ to the Hartree--Fock state. Such a state is prepared with $\hat S_{ijab}^{(2)} = \hat S_{ia}^{(1)}\hat S_{jb}^{(1)}$ applied to qubits $\mathcal S_V = \{i, j, a, b\}$ initialized to $\ket{0}$. The corresponding quantum circuit consists of two disentangled copies of $\hat S_{ia}^{(1)}$. The second type is generated by acting with
\[
    \frac{1}{\sqrt{3}}\Big(-\hat{E}_{jb}^{1,1}\hat{E}_{ia}^{1,-1} + \hat{E}_{jb}^{1,0}\hat{E}_{ia}^{1,0} - \hat{E}_{jb}^{1,-1}\hat{E}_{ia}^{1,1}\Big)
\]
on the Hartree--Fock state and is prepared with $\hat S_{ijab}^{(3)}$ applied to qubits $\mathcal S_V = \{i, j, a, b\}$ initialized to $\ket{0}$. We note that general circuits for preparing arbitrary CSFs can be obtained using recursive circuits \cite{Carbone2022} or Clebsch--Gordan transforms \cite{Bacon2006}. These constructions require controlled versions of arbitrary unitaries and are expensive in terms of nonlocal gates. Instead, we have designed state-specific circuits with minimal gate counts.

The rest of the state over the qubits $\mathcal Q \setminus \mathcal S_V$ is prepared by two unitaries, $\hat D_\mu$ and $\hat W_\mu^{(c)}$. Let $\mathcal Q^{(d)} \subseteq (\mathcal Q \setminus \mathcal S_V)$ denote the qubits corresponding to MOs that are doubly occupied prior to the action of the electron-pair excitations. In the compressed representation, qubits corresponding to unoccupied and doubly occupied MOs are in states $\ket{0}$ and $\ket{1}$, respectively, and are prepared by
\begin{equation}
    \hat D_\mu = \prod_{i \in \mathcal Q^{(d)}} \hat X_i
\end{equation}
acting on qubits initialized to $\ket{0}$. The unitary $\hat W_\mu^{(c)}$ consists of a sequence of compressed electron-pair rotations
\begin{equation}
    \hat U_{ia}(\theta) := e^{\theta \hat T_{ia}^{(c)}},
\end{equation}
generated by the compressed electron-pair excitation
\begin{equation}\label{eqn:tap_pair_exc}
    \hat T_{ia}^{(c)} = \frac{i}{2}\left(\hat X_i \hat Y_a - \hat Y_i \hat X_a\right)
\end{equation}
under the Jordan--Wigner mapping. The quantum circuit to perform each rotation $\hat U_{ia}(\theta)$, up to a global phase, is shown in \fig{fig:tap_pair_exc} \cite{Yordanov2021}.

\begin{figure}
    \centering
    \resizebox{\columnwidth}{!}{
    \begin{quantikz}
        \lstick{$\ket{0}$} && \gate{H}  & \octrl{1} & \ctrl{2} & \qw & \ctrl{1} & \rstick[2]{$\ket{\Phi_{\mu\nu}^{(\mathcal Q)}}$}\\
        \lstick{$\ket{0}$} & \qwbundle{n} & \qw & \gate{\hat S_\mu \hat D_\mu} & \qw & \gate{\hat W_{\mu}^{(c)}} & \swap{1} &\\
        \lstick{$\ket{0}$} & \qwbundle{n} & \qw & \qw & \gate{\hat S_\nu \hat D_\nu} & \gate{\hat W_{\nu}^{(c)}} & \targX{} &
    \end{quantikz}
    }
    \caption{Quantum circuit to prepare the states $\ket{\Phi_{\mu\nu}^{(\mathcal Q)}} = (\ket{0}\ket{\phi_\mu^{(\mathcal Q)}} + \ket{1}\ket{\phi_{\nu}^{(\mathcal Q)}})/\sqrt{2}$, where $\hat S_\mu \in \{\hat S^{(0)}, \hat S_{ia}^{(1)}, \hat S_{ijab}^{(2)}, \hat S_{ijab}^{(3)}\}$ prepares the initial compressed CSF state and $n = |\mathcal Q|$. The operators $\hat W_\mu^{(c)}$ are compressed electron-pair rotations. A CSWAP network, consisting of $n$ CSWAP operations, is added to obtain the state $\ket{\Phi_{\mu\nu}^{(\mathcal Q)}}$ on the first $n + 1$ qubits.}
    \label{fig:swap_test}
\end{figure}
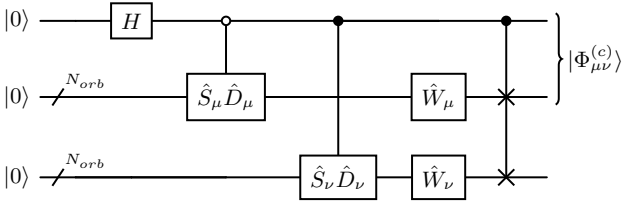

The $(|\mathcal Q| + 1)$-qubit state $\ket{\Phi_{\mu\nu}^{(\mathcal Q)}}$ is constructed using the quantum circuit in \fig{fig:swap_test} with $|\mathcal Q|$ ancilla qubits \cite{Baek2023}. This SWAP-test-based approach avoids adding controls to the pair-excitation rotations $\hat W_\mu^{(c)}$ because they act trivially on $\ket{0}^{\otimes |\mathcal Q|}$. Similarly, to obtain a controlled implementation of $\hat S_\mu \hat D_\mu$, only the single-qubit gates in the gate decomposition of $\hat S_\mu \hat D_\mu$ need to be promoted to their controlled versions, as the controlled gates within $\hat S_\mu \hat D_\mu$ also act trivially on $\ket{0}^{\otimes |\mathcal Q|}$.

\begin{figure}
    \centering
    \resizebox{\columnwidth}{!}{
    \subfloat[]{
    \centering
        \begin{quantikz}
            \lstick{$i$} & \gate{X} & \qw & \targ{} &\\
            \lstick{$a$} & \gate{X} & \gate{H} & \ctrl{-1} &
        \end{quantikz}
    }
    \subfloat[]{
    \centering
        \begin{quantikz}
            \lstick{$i$} & \gate{X} & \qw & \targ{} & \qw &\\
            \lstick{$j$} & \gate{X} & \qw & \qw & \targ{} &\\
            \lstick{$a$} & \gate{X} & \gate{H} & \ctrl{-2} & \qw &\\
            \lstick{$b$} & \gate{X} & \gate{H} & \qw & \ctrl{-2} &
        \end{quantikz}
    }
    }\\
    \subfloat[]{
    \centering
    \resizebox{\columnwidth}{!}{
        \begin{quantikz}
            \lstick{$i$} & \gate{H} & \qw & \qw & \ctrl{1} & \qw & \ctrl{2} & \ctrl{3} & \gate{X} & \\
            \lstick{$j$} & \gate{R_y(\theta_1)} & \gate{X} & \ctrl{1} & \targ{} & \qw & \qw & \qw & \qw &\\
            \lstick{$a$} & \gate{X} & \qw & \targ{} & \ctrl{1} & \targ{} & \targ{} & \qw & \qw &\\
            \lstick{$b$} & \qw & \qw & \qw & \gate{H} & \ctrl{-1} & \qw & \targ{} & \qw &
        \end{quantikz}
    }
    }
    \caption{Quantum circuits (a) $\hat S_{ia}^{(1)}$, (b) $\hat S_{ijab}^{(2)}$, and (c) $\hat S_{ijab}^{(3)}$ applied to qubits corresponding to singly occupied orbitals to obtain seniority-$\Omega = 2$ and $\Omega = 4$ CSF states. The rotation gate $R_y$ in $\hat S_{ijab}^{(3)}$ is parameterized by $\theta_1 = -2\tan^{-1}(1/\sqrt{2})$.}
    \label{fig:csf_circuits}
\end{figure}
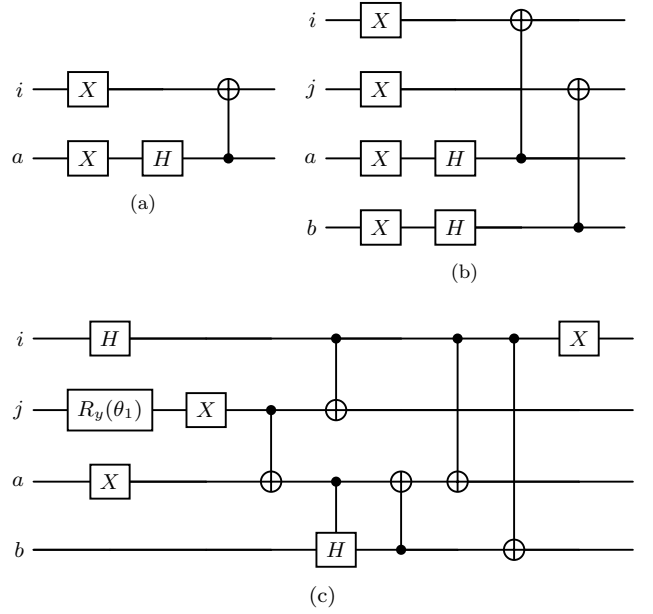

\begin{figure*}
    \centering
    \begin{quantikz}
        \lstick{i} & \gate{R^z(\frac{\pi}{2})} & \gate{R^x(\frac{\pi}{2})} & \ctrl{1} & \gate{R^x(\theta)} & \ctrl{1} & \gate{R^x(-\frac{\pi}{2})} & \gate{R^z(-\frac{\pi}{2})} &\\
        \lstick{a} & \gate{R^x(\frac{\pi}{2})} & & \targ{} & \gate{R^z(\theta)} & \targ{} & \gate{R^x(-\frac{\pi}{2})} & \qw &
    \end{quantikz}
    \caption{Quantum circuit implementing the rotation $\hat U_{ia}(\theta)$ generated by the compressed pair excitation in \eq{eqn:tap_pair_exc}.}
    \label{fig:tap_pair_exc}
\end{figure*}
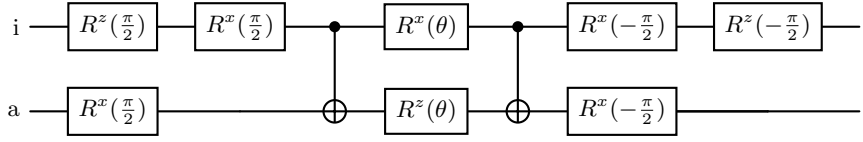

\subsection{Finite-Sampling Error Analysis}

Here, we derive a metric to quantify the sampling cost for energy estimation in Q-SENSE. For VQE, a common metric is the mean-square error of a single-shot estimator of the ground-state energy \cite{patelQuantumMeasurementQuantum2025}. This quantity cannot, in general, be evaluated exactly in quantum subspace methods, since the ground-state energy is not a simple function of the Hamiltonian matrix elements. Therefore, for Q-SENSE, we approximate the mean-square error using first-order perturbation theory. 

We first analyze finite-sampling error for the individual matrix elements $\mathsf{H}_{\mu\nu}$, which are estimated using the extended swap-test protocol [Eqs.~(\ref{swap_test_general}), (\ref{swap_test_state})] and a decomposition of the effective Hamiltonian into measurable fragments [Eq.~(\ref{partitioning})]. This setup is similar to single-energy estimation in a VQE experiment, for which the finite-sampling error analysis was carried out in Ref.~\cite{crawfordEfficientQuantumMeasurement2021}. Each fragment $\hat{F}_{\mu\nu}^{(\alpha)}$ has variance
\begin{equation}
    \sigma_{\mu\nu}^{(\alpha)2} = \braket{\Phi_{\mu\nu}|\hat{F}_{\mu\nu}^{(\alpha)2}|\Phi_{\mu\nu}} - \braket{\Phi_{\mu\nu}|\hat{F}_{\mu\nu}^{(\alpha)}|\Phi_{\mu\nu}}^2 .
\end{equation}
Given a total budget of $M_{\mu\nu}$ shots to estimate $\mathsf{H}_{\mu\nu}$, distributed among fragments as $M_{\mu\nu} = \sum_\alpha M_{\mu\nu}^{(\alpha)}$, the optimal allocation is found by minimizing the finite-sampling error using Lagrange multipliers. With this optimal allocation, the resulting variance of the estimator for $\mathsf{H}_{\mu\nu}$ is $\sigma_{\mu\nu}^2 / M_{\mu\nu}$, where
\begin{equation}
    \sigma_{\mu\nu} = \sum_\alpha \sigma_{\mu\nu}^{(\alpha)}. \label{matrix_element_variance}
\end{equation}

We now use perturbation theory to estimate the finite-sampling error in the ground-state energy. Let $\mathsf{S}$ denote an estimator of $\mathsf{H}$ that satisfies
\begin{equation}
    \mathsf{S} = \mathsf{H} + \mathsf{E},
\end{equation}
where $\mathsf{E}$ is an error matrix with Gaussian entries $\mathsf{E}_{\mu\nu}$ of mean $0$ and variance $\sigma_{\mu\nu}^2 / M_{\mu\nu}$, which are independent subject to the symmetry condition $\mathsf{E}_{\mu\nu} = \mathsf{E}_{\nu\mu}$. The mean-square error of our estimate of the ground-state energy of $\mathsf{H}$ is
\begin{equation}
    \epsilon^2 = \mathbb{E}\big[(E_\text{min}(\mathsf{S}) - E_\text{min}(\mathsf{H}))^2\big], \label{mse_generic}
\end{equation}
where $E_\text{min}$ denotes the map from a Hermitian matrix to its ground-state energy. The first-order correction $E^{(1)}$ to $E_\text{min}(\mathsf{S}) - E_\text{min}(\mathsf{H})$ is
\begin{equation}
    E^{(1)} = \vec{c}_0^{\ T} \mathsf{E} \vec{c}_0,
\end{equation}
where $\vec{c}_0$ is the ground-state eigenvector of the subspace Hamiltonian $\mathsf{H}$. A straightforward calculation gives the following first-order estimate of the mean-square error in \eq{mse_generic}:
\begin{align}
    \mathbb{E}\left[\big(E^{(1)}\big)^2\right] &= \Var\left(E^{(1)}\right)\nonumber\\
    &= \sum_\mu c_{0,\mu}^4 \frac{\sigma_{\mu\mu}^2}{M_{\mu\mu}} + 4\sum_{\mu < \nu} c_{0,\mu}^2 c_{0,\nu}^2 \frac{\sigma_{\mu\nu}^2}{M_{\mu\nu}}. \label{perturb_estimate_1storder}
\end{align}
Using the same Lagrange-multiplier approach as in Ref.~\cite{crawfordEfficientQuantumMeasurement2021}, we obtain the optimal allocation of measurements and, to first order, the relation between shot count and mean-square error for Q-SENSE:
\begin{equation}
    \epsilon^2_Q = \frac{1}{M}\left[\sum_\mu c_{0,\mu}^2 \sigma_{\mu\mu} + 2\sum_{\mu < \nu} |c_{0,\mu} c_{0,\nu}| \sigma_{\mu\nu}\right]^2. \label{sampling_cost_metric}
\end{equation}
The constant $\epsilon^2_Q M$, which is determined by the standard deviations of the matrix-element estimators and the Q-SENSE ground state, serves as a metric for the energy-estimation cost in Q-SENSE. 

\begin{figure}
    \centering\hspace*{-10pt}
    \includegraphics[width=0.9\columnwidth]{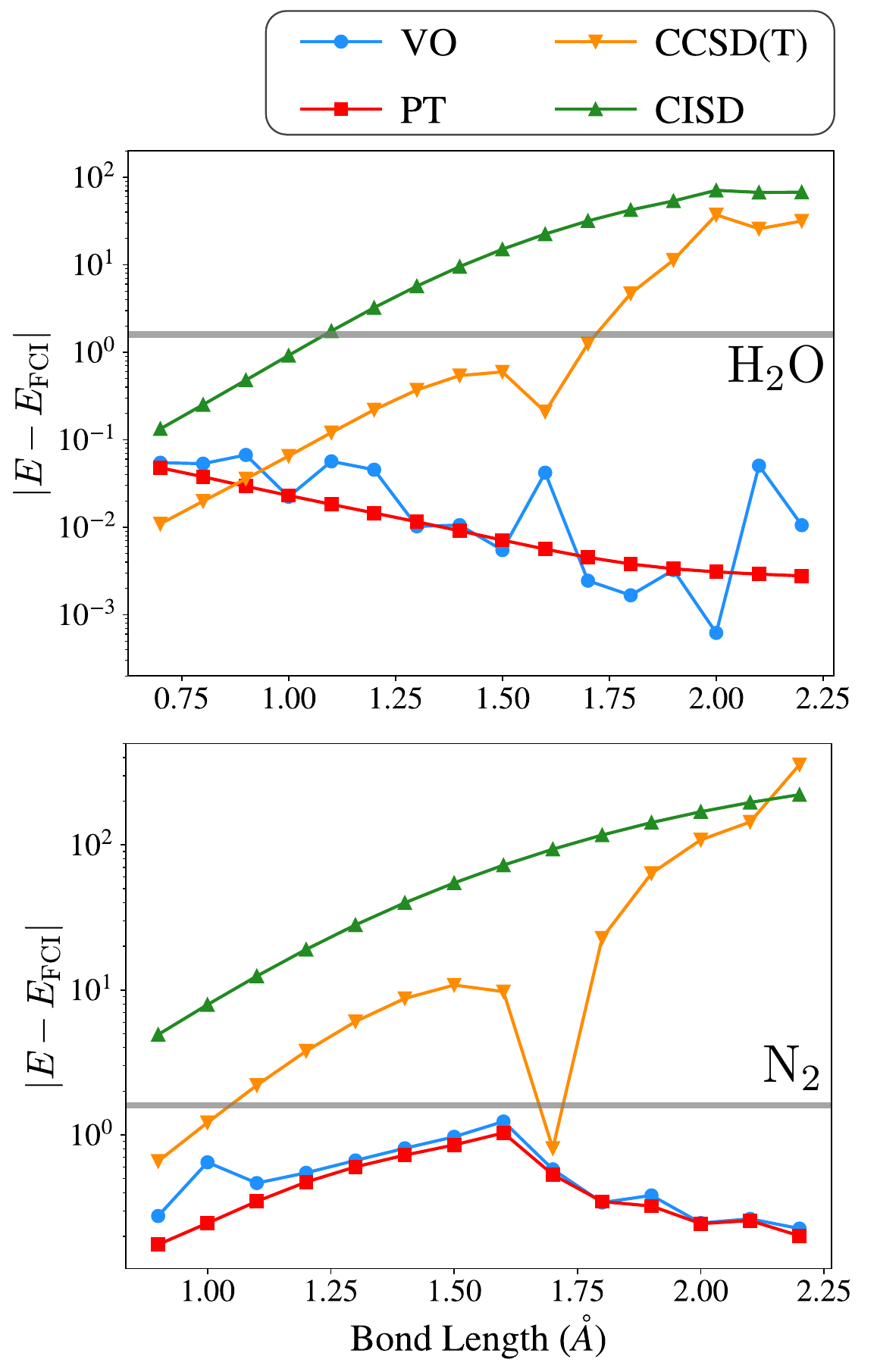}
    \caption{Errors relative to FCI for various bond lengths of H$_2$O and N$_2$. The horizontal gray line denotes chemical accuracy (1.6 mHa).}
    \label{fig:fci_error}
\end{figure}

\section{Results}

Here, we assess the accuracy and computational cost of the Q-SENSE framework on the molecular electronic Hamiltonians of H$_2$O (with $\angle \text{HOH} = 107.6^\circ$) and N$_2$ in the STO-3G basis set. To assess performance in both weakly and strongly correlated regimes, we consider stretching of the NN bond in N$_2$ and symmetric stretching of the OH bonds in H$_2$O. All Hamiltonians were generated with OpenFermion \cite{mccleanOpenFermionElectronicStructure2020} using PySCF \cite{sunPySCFPythonbasedSimulations2018,sunRecentDevelopmentsPySCF2020} as a backend, and the Jordan--Wigner transformation \cite{jordanUeberPaulischeAequivalenzverbot1928} was used to express the Hamiltonians in terms of Pauli operators. We use a full-valence active space to define $\mathcal{A}$, i.e., 6 orbitals and 8 electrons for H$_2$O and 8 orbitals and 10 electrons for N$_2$. The 1s core orbitals of O and N for H$_2$O and N$_2$, respectively, are in $\mathcal{I}$. For the STO-3G basis set and this selection of $\mathcal{A}$, $\mathcal{V}$ is the empty set. We benchmark the Q-SENSE framework on three metrics: (1) accuracy of the ground-state energies; (2) complexity of Q-SENSE basis-state preparation; and (3) sampling cost of estimating the Q-SENSE ground-state energy.

\subsection{Energies}

We first evaluate the performance of the Q-SENSE framework for symmetric bond stretching of H$_2$O. Figure~\ref{fig:fci_error} shows the error in the ground-state energy relative to the full configuration interaction (FCI) result for the VO and PT versions of Q-SENSE and for two classical methods: configuration interaction with singles and doubles (CISD) and coupled cluster with singles and doubles and perturbative triples (CCSD(T)). Both Q-SENSE variants successfully achieve chemical accuracy, defined as an FCI error $< 1.6$ mHa, along the dissociation curve, and are therefore viable in both weakly and strongly correlated regimes. The same success is observed for N$_2$ triple-bond stretching, which introduces significant strong correlation. The superiority of Q-SENSE over CISD and CCSD(T) is apparent, especially at the dissociation limits.

\subsection{Features of the Q-SENSE Basis}

We now analyze properties of the Q-SENSE basis states. As shown in Table~\ref{tab:circuit_empirical}, the VO method produces a much smaller basis than PT but requires more generators in the associated electron-pair-rotation unitaries. This demonstrates the flexibility of the Q-SENSE framework, which can trade off the size of the classical eigenvalue problem against the complexity of the quantum circuit required to implement $\hat{W}_{\mu}$. Despite this, both methods achieve chemical accuracy with a smaller basis than CISD, which involves 51 and 184 Slater determinants with nonzero coefficients for H$_2$O and N$_2$, respectively.

To quantify the quantum-circuit cost of the Q-SENSE method, we use the extended swap-test circuits in \fig{fig:swap_test} as representative. They are used to estimate off-diagonal matrix elements and are more resource-intensive than circuits for the diagonal elements. The circuit cost is quantified in terms of the total CNOT gate count and the circuit depth, which are the primary circuit-related bottlenecks on near-term hardware. To obtain resource estimates, we optimized the mapping of the circuits to CNOTs and single-qubit gates using Qiskit’s \texttt{transpile} function with \texttt{optimization\_level=3}, \texttt{basis\_gates=['u3', 'cx']} and all-to-all qubit connectivity \cite{javadi2024quantum}.

\begin{table}[]
\renewcommand{\arraystretch}{1.1}
\begin{tabular*}{\columnwidth}{@{\extracolsep{\fill}}ccccccccc}
    \toprule
    \multirow{2}{*}{Molecule} & \multirow{2}{*}{Method} & \multirow{2}{*}{$N_\text{states}$} & \multicolumn{2}{c}{$N_\text{pair}$} & \multicolumn{2}{c}{CNOTs} & \multicolumn{2}{c}{Depth} \\
    \cmidrule(lr){4-5} \cmidrule(lr){6-7} \cmidrule(lr){8-9}
    & & & Avg. & Max. & Avg. & Max. & Avg. & Max. \\
    \midrule
    \multirow{2}{*}{H$_2$O} & VO & 11 & 4 & 30 & 60 & 130 & 68 & 142 \\
    & PT & 37 & 1 & 2 & 24 & 63 & 30 & 77 \\
    \midrule
    \multirow{2}{*}{N$_2$} & VO & 23 & 9 & 63 & 98 & 261 & 96 & 219\\
    & PT & 166 & 4 & 6 & 23 & 101 & 24 & 105 \\
    \bottomrule
\end{tabular*}
\caption{Classical and quantum resource requirements for the VO and PT methods applied to H$_2$O and N$_2$ at 1.0~\AA\ bond length. Classical resources are measured by the subspace-basis size $N_\text{states}$. Quantum resources include the number of electron-pair rotations $N_\text{pair}$ in the $\hat{W}_{\mu}$ unitaries, and the CNOT counts and circuit depths for the state-preparation circuits used to estimate off-diagonal Hamiltonian matrix elements. For quantum resources, we report both average and maximum (worst-case) values across all matrix elements.}
\label{tab:circuit_empirical}
\end{table}

\begin{table}[]
    \centering
    \begin{tabular*}{\columnwidth}{@{\extracolsep{\fill}} lcc}
    \toprule
         Component & CNOT & Depth \\
         \midrule
         Controlled $\hat S^{(0)}\hat D_\mu$ & $N_e/2$ & $N_e/2$\\\midrule
         Controlled $\hat S_{ia}^{(1)}\hat D_\mu$ & $3 + N_e/2$ & $5 + N_e/2$\\\midrule
         Controlled $\hat S_{ijab}^{(2)}\hat D_\mu$ & $6 + N_e/2$ & $8 + N_e/2$\\\midrule
         Controlled $\hat S_{ijab}^{(3)}\hat D_\mu$ & $8 + N_e/2$ & $9 + N_e/2$\\\midrule
         $\hat U_{ia}(\theta)$ & 2 & 5\\\midrule
         CSWAP network & $7N_{\text{orb}}$ & $12N_{\text{orb}}$\\
         \bottomrule
    \end{tabular*}
    \caption{CNOT count and circuit depth of various components of the state-preparation circuits used to estimate off-diagonal matrix elements, where $N_{\rm orb}$ and $N_e$ are the numbers of orbitals and electrons, respectively.}
    \label{tab:circuit_comp}
\end{table}

Table~\ref{tab:circuit_comp} shows the CNOT counts and depths of the individual circuit components of the extended swap-test state-preparation circuit. The dependence of the CNOT count and depth on the number of electrons $N_e$ arises from the need to control state preparation on an ancilla qubit. For example, $(N_e - \Omega)/2$ CNOTs are required to prepare $\ket{1}$ states on the qubits corresponding to doubly occupied orbitals, conditioned on the ancilla qubit. Pair excitations can be implemented with a constant and relatively modest CNOT count and circuit depth. The CSWAP network required for the swap-test state introduces CNOT and depth overheads that scale with system size. Thus, the CNOT count and circuit depth scale linearly with the system size $N_e$ and with the number of generators in $\hat{W}_\mu$. The empirical quantum-circuit costs for the PT and VO routines are summarized in Table~\ref{tab:circuit_empirical} for both H$_2$O and N$_2$. The table presents both worst-case and average-case metrics calculated over the full set of matrix elements required for each method. Note that the reduction in circuit cost for PT relative to VO arises from two sources. First, PT uses fewer generators in $\hat{W}_\mu$ on average; second, PT typically admits larger classical sets $\mathcal{C}$ than quantum sets $\mathcal{Q}$ for matrix-element estimation.

\subsection{Quantum Measurements}

\begin{figure}
    \centering\hspace*{-25pt}
    \includegraphics[width=0.9\columnwidth]{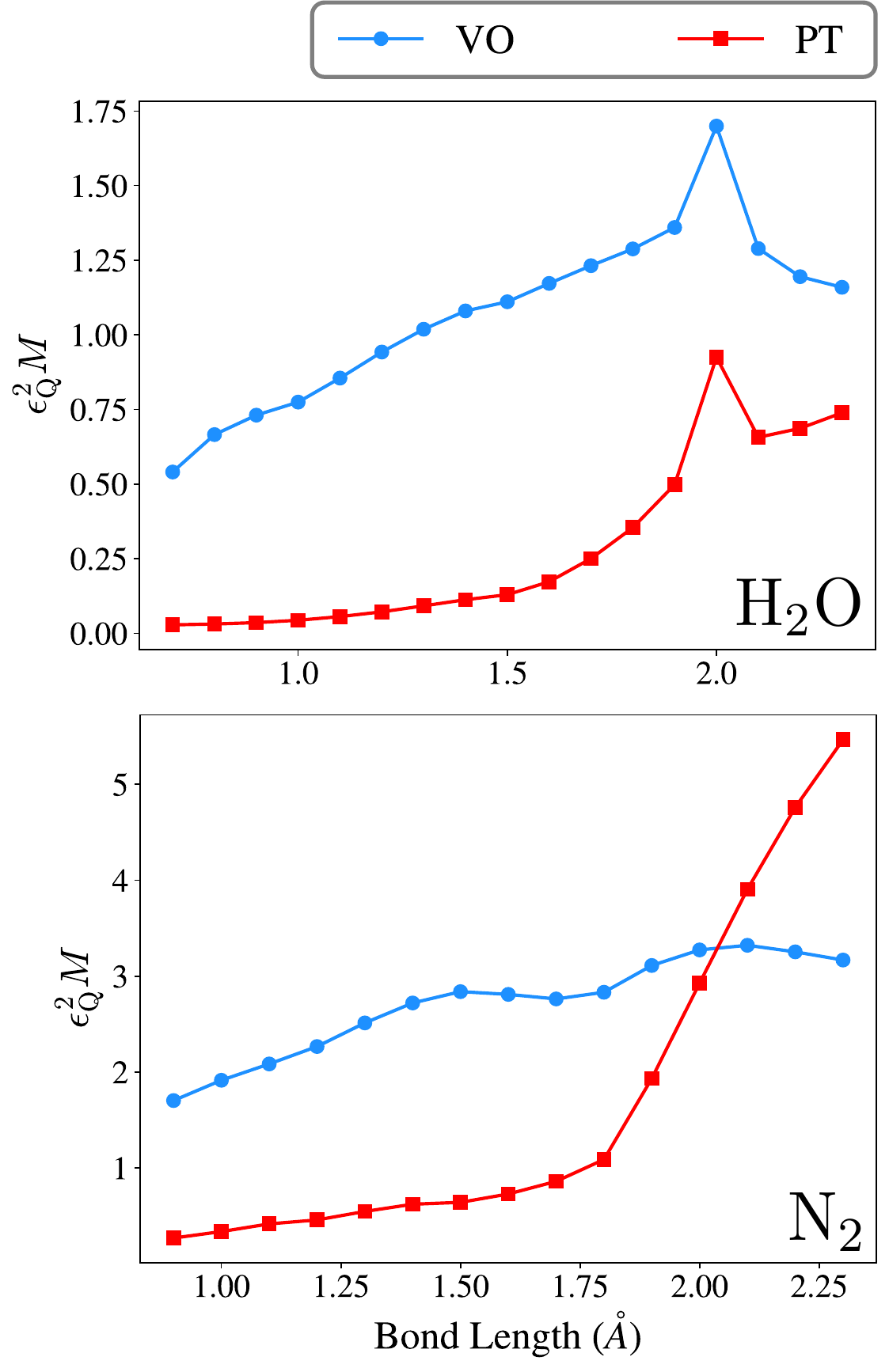}
    \caption{Sampling-cost metric $\epsilon^2_Q M$ for the VO and PT methods at various bond lengths of H$_2$O and N$_2$.}
    \label{fig:sampling_cost}
\end{figure}

We now analyze the measurement cost required to estimate the ground-state energy in the Q-SENSE subspace. Figure~\ref{fig:sampling_cost} presents the energy-estimation cost metric $\epsilon^2_Q M$ for both H$_2$O and N$_2$ across all bond lengths. The metric $\epsilon^2_Q M$ is the proportionality constant relating the number of shots and the mean-square error in the ground-state-energy estimator. Based on an estimated 350~$\mu$s per shot on an IBM quantum processor \cite{IBMQuantum2023}, achieving chemical accuracy for the most costly points on the H$_2$O and N$_2$ curves would require approximately 4 and 8 minutes for the VO method and 2 and 12 minutes for the PT method, respectively. Figure~\ref{fig:sampling_cost} also shows that PT usually has a lower energy-estimation cost than VO, even though it requires estimating more matrix elements overall. PT becomes more expensive than VO only at stretched geometries for N$_2$. On average, the PT basis states are “more classical,” since their $\hat{W}_\mu$ operators act on smaller sets of MOs. Consequently, a larger portion of matrix-element evaluations in the PT method are performed classically, quantified by larger $|\mathcal{C}|$ relative to $|\mathcal{Q}|$. This reduces the overall PT measurement cost despite the increased number of matrix elements. 

These observations illustrate how Q-SENSE interpolates between VQE and CI. PT operates closer to the CI limit, which has a sampling cost of zero, whereas VO operates closer to the VQE limit, which has $N_\text{state} = 1$. 

In Table~\ref{tab:vqe_comparison}, we compare the energy-estimation costs for VO, PT, and VQE for equilibrium, correlated, and dissociated geometries of both systems. The VQE results were taken from Ref.~\cite{choiProbingQuantumEfficiency2023}. Both Q-SENSE variants exhibit lower evaluation costs for all geometries. Moreover, PT circumvents two of the most expensive quantum-measurement bottlenecks of both VQE and VO: (i) evaluation of gradients of state-preparation circuit parameters and (ii) the large number of optimization loops, both of which require many additional energy estimations. This provides strong evidence that PT is the most efficient method from a measurement perspective. Despite this, PT involves a variational optimization of molecular orbitals [Eq.~(\ref{eq:orbital_opt})]. A warm start for this orbital optimization can be obtained using a multiconfiguration self-consistent field (MCSCF) calculation, which optimizes orbitals to minimize the ground-state energy in the space spanned by the initial CSFs $\{\hat V_\mu\ket{\rm HF}\}$, i.e., a Q-SENSE solution obtained without classically hard $\hat{W}_\mu$ operators.

Interestingly, the energy-estimation cost for Q-SENSE has a significant dependence on bond length, particularly for the PT variant. This trend, visible in Fig.~\ref{fig:sampling_cost}, contrasts with the relatively constant VQE costs for these systems at different geometries \cite{choiProbingQuantumEfficiency2023} (see Table~\ref{tab:vqe_comparison}). This sensitivity of measurement cost to molecular geometry highlights a feature of Q-SENSE, and of quantum subspace methods more generally: by sampling components of the wave function rather than the whole, they can allocate samples to the most important parts—a capability that VQE lacks. In our cost analysis, this feature manifests through the presence of the ground-state eigenvector coefficients in the sampling-cost metric [\eq{sampling_cost_metric}], whose magnitudes quantify the significance of the associated matrix elements for estimating the ground-state energy. At strongly correlated geometries (e.g., stretched N$_2$), the overhead increases because a larger number of Q-SENSE basis states contribute nontrivially to the electronic ground state. Consequently, the PT variant, which spreads the wave function over more Q-SENSE basis states, shows a more pronounced dependence of energy-estimation cost on bond length. This adaptive allocation of resources to resolve the most relevant parts of the wave function is conceptually similar to the classically boosted VQE (CB-VQE) method \cite{radinClassicallyBoostedVariationalQuantum2021a}, which splits the wave function into classical and quantum parts. In CB-VQE, the classical and quantum components are not orthogonal, whereas Q-SENSE basis states are. Thus, Q-SENSE measurements are even more efficient than those in CB-VQE.     

Note that our analysis is based on certain idealizations. The cost estimates do not account for the effects of hardware noise, which would increase the required number of measurements. Furthermore, our optimal measurement-allocation scheme assumes prior knowledge of the ground-state eigenvector and variances, which in practice would need to be estimated either classically or via a small number of initial samples. Despite these approximations, the results demonstrate how Q-SENSE reduces the measurement cost compared with VQE.

\begin{table}[t!]
\centering
\begin{tabular*}{\columnwidth}{@{\extracolsep{\fill}} lcccc}
\toprule
System                  & Bond Length (\AA) & VQE  & VO                 & PT                    \\
\midrule
\multirow{3}{*}{H$_2$O} & 1.0               & 7.87 & 0.66               & $3 \times 10^{-2}$  \\
                        & 2.1               & 9.08 & 1.36               & 0.50                  \\
                        & 3.0               & 0.66 & $2 \times 10^{-4}$ & $3 \times 10^{-4}$ \\
\midrule
\multirow{3}{*}{N$_2$}  & 1.2               & 9.57 & 2.27               & 0.46                  \\
                        & 1.4               & 11.6 & 2.72               & 0.62                  \\
                        & 2.2               & 5.21 & 3.25               & 4.76                  \\
\bottomrule
\end{tabular*}
\caption{Comparison of sampling-cost metrics for a single evaluation of the ground-state energy in VQE, VO, and PT for H$_2$O and N$_2$ at various bond lengths. VQE results are taken from Ref.~\cite{choiProbingQuantumEfficiency2023}. All results were obtained using the FC-SI decomposition.}
\label{tab:vqe_comparison}
\end{table}

\begin{table}[t]
\renewcommand{\arraystretch}{1.1}
\begin{tabular*}{\columnwidth}{@{\extracolsep{\fill}}lcccc}
    \toprule
    \multirow{2}{*}{Molecule} & \multicolumn{2}{c}{$N_\text{terms}$} & \multicolumn{2}{c}{$\lambda$} \\
    \cmidrule(lr){2-3} \cmidrule(lr){4-5}
           & Avg.     & Max.     & Avg.   & Max. \\
    \midrule
    H$_2$O & 0.03     & 0.06     & 0.06   & 0.76\\
    N$_2$  & 0.01     & 0.06     & 0.05   & 0.81\\
    \bottomrule
\end{tabular*}
\caption{Reduction in Hamiltonian complexity after classical partial matrix-element evaluation using the $\{\mathcal{C}, \mathcal{Q}\}$ partition, for H$_2$O and N$_2$ at 1.0~\AA\ bond length. The values represent the ratios of the number of Pauli terms ($N_\text{terms}$) and the 1-norm ($\lambda$) in the effective Hamiltonians $\hat{h}_{\mu\nu}$ to those in the original Hamiltonian $\hat{H}$. Average and maximum (worst-case) ratios are shown, calculated over the set of matrix elements for the VO method. The constant-term optimization technique (Sec.~\ref{sec:constant}), which can change the 1-norm of the effective Hamiltonians $\hat{h}_{\mu\nu}$ for off-diagonal matrix elements within the extended swap-test formalism, was not used when calculating the 1-norms.}
\label{tab:tapering_effect}
\end{table}

We now consider the role of classical partial matrix-element evaluation, obtained via the $\{\mathcal{C}, \mathcal{Q}\}$ partition described in Sec.~\ref{sec:estimation_general}, in reducing the energy-estimation cost. Table~\ref{tab:tapering_effect} shows the complexity of the effective Hamiltonians after partial matrix-element evaluation, compared with the original molecular electronic Hamiltonian. We quantify this using the number of Pauli terms $N_\text{terms}$ and the 1-norm $\lambda$ of the Pauli coefficients, which provides an upper bound on the variance of the Hamiltonian for any quantum state \cite{patelQuantumMeasurementQuantum2025}. For both H$_2$O and N$_2$, partial matrix-element evaluation reduces the number of Pauli terms by over 93\% across all matrix elements. The data also show a larger reduction in the number of Pauli terms on average for N$_2$ than for H$_2$O. This trend is expected, since the probability of removing a term via classical partial matrix-element evaluation scales exponentially with the number of orbitals $N_\text{orb}$, whereas the total number of Hamiltonian terms grows only polynomially. However, this improvement for individual matrix elements is offset by the larger number of simultaneous eigen-subspaces of the orbital-seniority operators that one must consider as system size increases. The average reduction in the 1-norm is also substantial, exceeding 94\% for both systems, with larger average improvements observed for N$_2$ compared with H$_2$O. However, the worst-case reduction across all matrix elements is much more modest. For both molecules, this occurs when the bra and ket states are in the seniority-zero sector. This is expected, since the seniority-zero sector contributes most strongly to the ground state, as reflected in the large 1-norm of its Hamiltonian terms.

\section{Conclusions}

We have introduced and benchmarked the Q-SENSE framework as a scalable alternative to traditional near-term quantum algorithms for molecular electronic structure. Unlike adaptive variational approaches, whose circuit depth typically grows rapidly with system size, Q-SENSE avoids this bottleneck by trading off circuit complexity against the number of basis functions. This balance allows the method to interpolate between two well-known limits: a single basis function reproduces VQE, while a complete classical representation recovers configuration interaction (CI).

A key feature of Q-SENSE is its use of basis functions constructed from eigenstates of orbital-seniority operators. These states are orthogonal by construction, avoiding the generalized-eigenvalue problem that often leads to ill-conditioned matrices and numerical instabilities in other quantum subspace methods. The seniority structure also enables reductions in measurement cost that generalize qubit tapering, contributing to the overall efficiency of the approach.

Our numerical results on H$_2$O and N$_2$ demonstrate that Q-SENSE achieves chemical accuracy across both weakly and strongly correlated regimes, including challenging bond-stretching dissociation curves. The VO and PT variants each offer complementary trade-offs: VO uses a compact basis at the expense of more complex unitaries that require parameter optimization, whereas PT employs a larger basis but simpler circuits that are exempt from parameter optimization. In both cases, chemical accuracy is achieved with fewer basis states than CISD, and measurement costs are substantially reduced relative to VQE. These findings indicate that Q-SENSE is well suited to capture strong-correlation effects while remaining feasible for near-term quantum hardware.

Conceptually, Q-SENSE can be viewed as a quantum extension of the MCSCF method. In addition to Slater-determinant coefficients and orbital optimization familiar from MCSCF, Q-SENSE incorporates pair-rotation operations that conserve seniority. While these rotations do not form a polynomial-size Lie group, they can nevertheless be efficiently represented by quantum circuits. This property, together with the measurement advantages of seniority symmetry, provides Q-SENSE with a distinctive balance of accuracy and efficiency.

Overall, Q-SENSE eliminates the need for nonlinear optimization over circuit parameters in the PT variant and significantly reduces measurement overhead compared with traditional VQE in both the PT and VO variants. Our benchmarks on small molecules show that the method is compatible with current quantum-hardware assumptions, with sampling times in the range of minutes on today’s superconducting quantum processors. These results highlight Q-SENSE as a promising route toward scalable, symmetry-exploiting quantum algorithms for electronic structure in both weakly and strongly correlated regimes.


\section*{Acknowledgments}

We would like to thank Ilya Ryabinkin for useful discussions and Ruben Van der Stichelen for helpful comments on the manuscript.
S.P., T.Z., and A.F.I. acknowledge financial support from the Natural Sciences 
and Engineering Council of Canada (NSERC). This research was partly enabled by the support of Compute Ontario (computeontario.ca) and the Digital Research Alliance of Canada (alliancecan.ca). Part of the computations were performed on the Niagara and Trillium supercomputers at the SciNet HPC Consortium, and the NARVAL and RORQUAL supercomputers under the Calcul Quebec Consortium. SciNet is funded by Innovation, Science, and Economic Development Canada, the Digital Research Alliance of Canada, the Ontario Research Fund: Research Excellence, and the University of Toronto. 

\bibliography{library}

\end{document}